\documentclass[article]{article}
\usepackage{makeidx,epsfig}
\usepackage{setspace,graphicx}
\usepackage[margin=1in]{geometry}
\usepackage{algorithm2e}
\usepackage{amsmath}

\def\ket#1{|{#1}\rangle}

\def\braket#1#2{\langle{#1}|{#2}\rangle}

\newcommand{\be}{\begin{equation}}
\newcommand{\ee}{\end{equation}}
\newcommand{\ba}{\begin{eqnarray}}
\newcommand{\ea}{\end{eqnarray}}
\newcommand{\ban}{\begin{eqnarray*}}
\newcommand{\ean}{\end{eqnarray*}}

\usepackage[usenames,dvipsnames]{color}

\begin{document}
\title{The Quantum Frontier}
\author{Joseph F. Fitzsimons 
\\Centre for Quantum Technologies \\National University of Singapore \and 
Eleanor G. Rieffel \\ NASA Ames Research Center \and
Valerio Scarani
\\Centre for Quantum Technologies \\National University of Singapore 
}
\maketitle

\section{Summary}

The success of the abstract model of computation, in terms of bits, logical 
operations, programming language constructs, and the like, makes it easy to
forget that computation is a physical process. Our cherished notions of
computation and information are grounded in classical mechanics, but the
physics underlying our world is quantum. In the early 80s researchers
began to ask how computation would change if we adopted a quantum
mechanical, instead of a classical mechanical, view of computation. 
Slowly, a new picture of computation arose, one that gave rise to
a variety of faster algorithms, novel cryptographic mechanisms,
and alternative methods of communication. Small quantum information
processing devices have been built, and efforts are underway to build 
larger ones. Even apart from the existence of these devices, the
quantum view on information processing has provided significant
insight into the nature of computation and information, and a deeper
understanding of the physics of our universe and its connections with
computation. 

We start by describing aspects of quantum mechanics that are at the
heart of a quantum view of information processing. We give our own
idiosyncratic view of a number of these topics in the hopes of
correcting common misconceptions and highlighting aspects that are often 
overlooked. A number of the phenomena described were initially viewed
as oddities of quantum mechanics, whose meaning was best left to
philosophers, topics that respectable physicists would avoid or, at
best, talk about only over a late night beer. It was quantum information
processing, first quantum cryptography and then, more dramatically,
quantum computing, that turned the tables and showed that these
oddities could be put to practical effect. It is these applications
we describe next. We conclude with a section describing some of the
many questions left for future work, especially the mysteries surrounding
where the power of quantum information ultimately comes from.

\section{A message in a quantum (or two)}

We begin a deeper excursion into the quantum view of information
processing by examining the role of randomness in quantum mechanics,
how that role differs from its role in classical mechanics, and
the evidence for the deeper role of randomness in quantum mechanics.

\subsection{Intrinsic randomness}

Randomness, or unpredictability, has been accepted by most human beings throughout history, based on the simple observation that events happen that nobody had been able to predict. The positivist program \cite{Comte1830,Mill1865} was a daring attempt of getting rid of randomness: as Laplace famously put it, everything would be predictable for a being capable of assessing the actual values of all physical degrees of freedom at a given time. This ultimate physical determinism is protected against the trivial objection that some phenomena remain unpredictable in practice, even in the age of supercomputers. Indeed, such phenomena involve too many degrees of freedom, or require too precise a knowledge of some values, to be predictable with our means. To put it differently, randomness appears as \textit{relative} to our knowledge and computational power: what is effectively random today may become predictable in the future.

Humanity being so familiar with randomness and science having apparently tamed it, the statement that \textit{quantum phenomena entail an element of randomness} hardly stirs any emotion. The trained scientists translate it as ``quantum physics deals with complex phenomena", the others as ``I have a good excuse not to understand what these people are speaking about". However, what nature is telling us through quantum physics is different: quantum phenomena suggest that there is \textit{intrinsic} randomness in our universe. In other words, some events are unpredictable even for Laplace's being, who knows all that can be known about this physical universe at a given time. It is \textit{absolute} randomness. Coming from scientific quarters, this claim sounds even more daring than positivism. What is the basis for it?

Physicists may well reply that quantum randomness is like the heart in the 
cup of coffee or the rabbit in the moon: once you have seen it, you see it 
always. For more convincing evidence, they can point to an outstanding 
phenomenon: the \textit{observation of the violation of Bell's inequalities}.

\subsection{Violation of Bell's inequalities\label{sec:Bell}}

In order to appreciate the power of Bell's inequalities \cite{Bell-64}, 
some notions need to be introduced. In particular, we must be precise
about what measurement means, whether in a quantum or classical setting.
Once the concepts around measurement are clear, Bell's inequalities
follow from a simple statistical argument about probabilities of
measurement outcomes.

\subsubsection{Description of the measurement process}

It is crucial to begin by providing an 
\textit{operational description of a measurement process}, which can 
be decomposed into three steps:
\begin{enumerate}
\item First, a physical system to be measured must enter the measurement 
device. Here we do not need to know anything about the physical system 
itself: we just assume that something indicates when the device is 
``loaded" (actual experiments do not even need this heralding step, but 
we assume it here for simplicity of the discussion).
\item The device has a knob, or some similar method of selecting settings, with
each position of the knob corresponding to a different measurement: 
the \textit{input} part of the measurement process consists 
of \textit{choosing a setting}, that is, a position of the knob. 
Two remarks must be made on this step. First, it is assumed 
that the process that chooses the setting 
is uncorrelated from the system to be measured. In short, we say that 
the choice is ``free." Second, it is \textit{not} assumed that different 
settings correspond to different measurements within the device: 
\textit{a priori}, the position of the knob may be uncorrelated with 
the physical measurement that is actually performed. We are going to sort 
the results conditioned on the setting, which is the only information 
about the input we can access.

\item The \textit{output} of a measurement process is readable information. 
Think of one lamp out of several being lit. For simplicity, in this paper 
we consider the example of binary information: the output of a measurement 
can be either of two outcomes. Outputs are often \textit{labeled} by 
numbers for convenience: so, one may associate one lamp with ``0" and 
the other with ``1"; or alternatively, with ``+1" and ``-1" respectively. 
But this label is purely conventional and the conclusions should not 
depend crucially on it.
\end{enumerate}

With respect to the ``free'' choice of the setting in point 2,
bear in mind that one is not requiring the choice to be made by an 
agent supposedly endowed with ``free will". As will become clear later in this section, what we need is that for two
different measuring systems, each consisting of a measuring device and
and object being measured, the measurement outcome of system $A$ cannot
depend on the setting of system $B$, and vice versa. If system $A$
has no knowledge of system $B$'s setting, and vice versa, there can be
no such dependence. It is impossible to rule out the possibility
of such knowledge completely, but steps can be taken to make it
appear unlikely. First, communication between the two systems can be 
ruled out by placing the systems sufficiently far apart that even 
light cannot travel between them during the duration of the experiment,
from measurement setting to reading the measurement outcome. Even without
communication, the outcome at system $A$ could depend on the setting
at system $B$ if the setting is predictable. In actual experiments, 
the choice is made by a physical random process, which is very 
reasonably assumed to be independent of the quantum systems to be measured. 
We must be clear about what sort of randomness is required since we will
use such a setup to argue for intrinsic randomness. We do not need
an intrinsically random process, just a process for choosing a setting
for system $A$ that is reasonably believed to be unpredictable to system $B$
and vice versa. A classical coin flip suffices
here, for example, even though the outcome is deterministic given the
initial position and momentum, because it is reasonable, though not
provable, that system $B$ does not have access to the outcome of a coin
flip at system $A$ and vice versa.

\subsubsection{Measurement on a single system}

Consider first the characterization of the results of a \textit{single} measurement device. The elementary measurement run (i.e. the sequence ``choose a setting -- register the outcome") is repeated many times, so that the statistics of the outcomes can be drawn. One observes, for instance, that for setting $x=1$ it holds $[\textrm{Prob}(0|x=1),\textrm{Prob}(1|x=1)]=[1/2,1/2]$; for setting $x=2$ it holds $[\textrm{Prob}(0|x=2),\textrm{Prob}(1|x=2)]=[1/3,2/3]$; for setting $x=3$ it holds $[\textrm{Prob}(0|x=3),\textrm{Prob}(1|x=3)]=[0.99,0.01]$; and so on for as many positions as the knob has. Apart from the recognition \textit{a posteriori} that some positions of the knob do correspond to something different happening in the device, what physics can we learn from this brute observation? Nothing much, and certainly \textit{not} the existence of intrinsic randomness. Indeed, for instance, setting $x=1$ may be associated to the instructions ``don't measure any physical property, just choose the outcome by tossing an unbiased coin". This counter-example shows that classical apparent randomness can be the origin of the probabilistic behavior of setting $x=1$. A similar argument can be made for settings $x=2$ and $x=3$, using biased coins.

\subsubsection{Measurement on two separate systems}
\label{sec:measOn2Systems}

However, things change dramatically if we consider \textit{two} measurement devices, if one further assumes that they cannot communicate (and there may be strong reasons to believe this assumption; ultimately, one can put them so far apart that not even light could propagate from one to the other during a measurement run). Then \textit{not all} statistical observations can be deconstructed with two classical processes as we did before. This is the crucial argument, so let us go carefully through it.

We denote by $x$ the input, the setting of the device at location A, 
and $a$ its outcome; $y$ the input of the device at location B, and $b$ 
its outcome. Moreover, $x$ and $y$ are assumed to be chosen independently 
of each other, so that the setting at A is unknown and unpredictable to 
location B, and vice versa. We restrict to the simplest situation, in 
which the choice of inputs is binary: so, from now on, in this section, 
$x,y\in\{0,1\}$.

First, let us discuss an example of a nontrivial situation, which can 
nevertheless be explained by classical pseudo-randomness. Suppose that 
one observes the following statistics for the probabilities 
$\textrm{Prob}(a,b|x,y)$ of seeing outcomes $a$ and $b$ given settings
$x$ and $y$:
\ban
\begin{array}{llll}
\textrm{Prob}(0,0|0,0)=1/2 & \textrm{Prob}(0,1|0,0)=0 & \textrm{Prob}(1,0|0,0)=0 & \textrm{Prob}(1,1|0,0)=1/2\\
\textrm{Prob}(0,0|0,1)=1/4 & \textrm{Prob}(0,1|0,1)=1/4 & \textrm{Prob}(1,0|0,1)=1/4 & \textrm{Prob}(1,1|0,1)=1/4\\
\textrm{Prob}(0,0|1,0)=1/4 & \textrm{Prob}(0,1|1,0)=1/4 & \textrm{Prob}(1,0|1,0)=1/4 & \textrm{Prob}(1,1|1,0)=1/4\\
\textrm{Prob}(0,0|1,1)=1/2 & \textrm{Prob}(0,1|1,1)=0 & \textrm{Prob}(1,0|1,1)=0 & \textrm{Prob}(1,1|1,1)=1/2\\
\end{array}\,
\ean In words, this means that $a=b$ when $x=y$, while $a$ and $b$ are 
uncorrelated when $x\neq y$. The presence of correlations indicate that 
uncorrelated coins are not a possible explanation. A classical explanation 
is possible, however. Assume that, in each run, the physical system to be 
measured in location A carries an instruction specifying that the value of 
the output should be $a_x$ if the setting is $x$; and similarly for 
what happens at B. In other words, from a common source, the physical
system sent to A receives instructions to answer $a_0$ if the setting is $0$
and $a_1$ if the setting is $1$ and the system sent to B receives 
instructions to answer $b_0$ if the setting is $0$ and $b_1$ if the 
setting is $1$. These instructions are summarized as
$\lambda=(a_0,a_1;b_0,b_1)$. The observation statistics above
require that the source emit only instructions
$\lambda=(a_0,a_1;b_0,b_1)$ such that $a_0=b_0$ and $a_1=b_1$. 
The precise statistics are obtained when the source chooses each 
of the four possible $\lambda$'s, $(0,0;0,0)$, $(0,1;0,1)$, 
$(1,0;1,0)$ and $(1,1;1,1)$, with equal probability, by coin flipping. 

Such a strategy is commonly referred to as \textit{pre-established agreement}; 
the physics jargon has coined the unfortunate name of 
\textit{local hidden variables} to refer to the $\lambda$'s. 
Whether a pre-established agreement can explain a table of probabilities
is only interesting if we assume that the output at A cannot depend on the
setting at B and vice versa. Without that requirement, any table of 
probabilities can be obtained by sending out the $16$ possible instructions
with the probabilities given in the table. This remark illustrates
why we insisted that the choice of setting must be made freely and
unpredictably. The ``local'' in ``local hidden variables'' refers
to the requirement that one side does not know the setting on the other side.
Before turning to the next example, it is important to stress a point: 
we are not saying that observed classical correlations \textit{must} 
be attributed to pre-established agreement (one can observe classical 
correlations by measuring quantum systems), rather that because classical 
correlations \textit{can} be attributed to pre-established agreement, it is 
impossible to use them to provide evidence for the existence of intrinsic 
randomness.

In order to finally provide such evidence, we consider the following 
statistics:
\ban
\begin{array}{llll}
\textrm{Prob}(0,0|0,0)=1/2 & \textrm{Prob}(0,1|0,0)=0 & \textrm{Prob}(1,0|0,0)=0 & \textrm{Prob}(1,1|0,0)=1/2\\
\textrm{Prob}(0,0|0,1)=1/2 & \textrm{Prob}(0,1|0,1)=0 & \textrm{Prob}(1,0|0,1)=0 & \textrm{Prob}(1,1|0,1)=1/2\\
\textrm{Prob}(0,0|1,0)=1/2 & \textrm{Prob}(0,1|1,0)=0 & \textrm{Prob}(1,0|1,0)=0 & \textrm{Prob}(1,1|1,0)=1/2\\
\textrm{Prob}(0,0|1,1)=0 & \textrm{Prob}(0,1|1,1)=1/2 & \textrm{Prob}(1,0|1,1)=1/2 & \textrm{Prob}(1,1|1,1)=0\\
\end{array}\,
\ean In words, it says that the $a=b$ for three out of four choices of 
settings, while $a\neq b$ for the fourth. Pre-established agreement cannot 
reproduce this table: it would require the fulfillment of the contradictory 
set of conditions $a_0=b_0$, $a_0=b_1$, $a_1=b_0$ and $a_1\neq b_1$. 
But consider carefully what this means: the outcomes of the measurement 
process cannot be the result of reading out a pre-existing list 
$\lambda=(a_0,a_1;b_0,b_1)$. Turn the phrase again and we are there: 
there was an element of unpredictability in the result of the measurements 
--- because, if all the results had been predictable, we could have listed 
these predictions on a piece of paper; but such a list cannot be written.

We have reached the conclusion that the observation of these statistics 
implies that the underlying process possesses intrinsic randomness. 
It is absolutely remarkable that such a conclusion can in principle be 
reached in a black-box scenario, as a consequence of observed statistics, 
without any discussion of the physics of the process.

One may further guess that the same conclusion can be reached if the 
statistics are not exactly the ones written above, but are not too far 
from those. This is indeed the case: all the statistics that can be 
generated by shared randomness must obey a set of linear inequalities; 
the statistics that violate at least one of those inequalities can be 
used to deduce the existence of intrinsic randomness. These are the 
famous \textit{Bell's inequalities}, named after John Bell who first 
applied these ideas to quantum statistics\footnote{As a curiosity, Boole had 
already listed many such inequalities \cite{Boole1862,Pitowsky94}, 
presenting them as conditions that must be trivially obeyed --- he could 
not expect quantum physics!}.

As a matter of fact, the statistics just described cannot be produced by measuring composite quantum systems at a distance: even in the field of randomness, there are some things that mathematics can conceive but physics cannot do for you. Nevertheless, quantum physics can produce statistics with a similar structure, in which $1/2$ is replaced by $p=(1+1/\sqrt{2})/4\approx 0.43$ and 0 by $1/2-p=(1-1/\sqrt{2})/4\approx 0.07$. This quantum realization still violates Bell's inequalities by a comfortable margin: one would have to go down to $p=3/8$, for the statistics of this family to be achievable with shared randomness (see Chapter 5 of \cite{sixpieces}). The bottom line of it all is: \textit{quantum physics violates Bell's inequalities, therefore there is intrinsic randomness in our universe}\footnote{Has nature disproved determinism? This is strictly speaking impossible. Indeed, full determinism is impossible to falsify: one may believe that everything was fully determined by the big bang, so that, among many other things, human beings were programmed to discover quantum physics and thus believe in intrinsic randomness. Others may believe that we are just characters in a computer game played by superior beings, quantum physics being the setting they have chosen to have fun with us (this is not post-modernism: William of Ockham, so often invoked as a paragon of the scientific mindset, held such views in the fifteen century). The so-called ``many-worlds interpretation" of quantum physics saves determinism in yet another way: in short, by multiplying the universes (or the branches of reality) such that all possibilities allowed by quantum physics do happen in this multiverse.

Hence, it is still possible to uphold determinism as ultimate truth; only, the power of Laplace's being must be suitably enhanced: it should have access to the real code of the universe, or to the superior beings that are playing with us, or to all the branches of reality. If we human beings are not supposed to have such a power, the observed violation of Bell's inequalities means, at the very least, that some randomness is ``absolute for us". As we wrote in the main text, but now with emphasis: there is intrinsic randomness in \textit{our} universe.}.

\subsection{Quantum certainties}
\label{sec:fewerWorlds}

While randomness is at the heart of quantum mechanics, it does not rule out certainty. Nor is randomness the whole story of the 
surprise of quantum mechanics: part of the surprise is that certain things that classical physics predicts should happen with certainty, quantum mechanics predicts do not happen, also with certainty. The most famous such example is provided by the Greenberger-Horne-Zeilinger correlations \cite{GHZ-90,GHZ-89,Pan-00}. These involve the measurement of three separated quantum systems. Given a set of observations, classical physics gives a clear prediction for another measurement: four outcomes can happen, each with equal probability, while the other four never happen. Quantum mechanics predicts, and experiments confirm, that exactly the opposite is the case (see Chapter 6 of \cite{sixpieces}).

So, quantum physics is not brute randomness. By making this observation, we take issue with a misconception
common in popular accounts of quantum mechanics, and some scholarly articles: 
that quantum mechanics, at least in the ``many-worlds interpretation," implies that every conceivable event happens in some universe\footnote{A typical quote: 
``There are even universes in which a given object in our universe has no 
counterpart - including universes in which I was never born and you 
wrote this article instead." \cite{Deutsch-98}}. On the contrary, as we 
just saw, there are conceivable possibilities (and even ones that a 
classical bias would call necessities) that cannot happen because 
of quantum physics. One of us has used this evidence to call for 
a ``fewer-worlds-than-we-might-think" interpretation of quantum 
mechanics \cite{Rieffel-07}.

\subsection{Think positive}

Because of other quantum properties, such as the inability to precisely 
measure both position and momentum (see Section \ref{sec:uncertaintyPrin}),
intrinsic randomness was rapidly accepted as the orthodox interpretation 
of quantum phenomena, four decades before the violation of Bell's 
inequalities was predicted (let alone observed). The dissenting voices, 
be they as loud as Einstein, Schr\"odinger and De Broglie, were 
basically silenced.

Even so, for more than half a century, physicists seemed to have 
succumbed to unconscious collective shame when it came to these matters. One perceives an underlying dejection in generations of physicists, and great physicists at that, who ended up associating quantum physics with insurmountable limitations to previous dreams of knowledge of and control over nature. The discourse was something along the lines of: ``You can't know position and momentum, because that's how it is and don't ask me why; now, shut up and calculate, your numbers will be very predictive of the few things that we can speak about". Several otherwise excellent manuals are still pervaded by this spirit.

It took a few people trained in both information science and quantum physics to realize that intrinsic randomness is not that bad after all: in fact, it is a \textit{very useful resource for some tasks}. And this is exactly the new, positive attitude: given that our universe is as it is, maybe we can stop complaining and try to do something with it. Moreover, since quantum physics encompasses classical physics and is broader than the latter, surely there must be tasks that are impossible with classical degrees of freedom, which become possible if one moves to quantum ones. Within a few years, this new attitude had triggered the 
entire field of \textit{quantum information science}.

The epic of the beginning of quantum information science have been told 
many times and its heroes duly sung. There is Wiesner dreaming of quantum 
money in the late 1970s and not being taken seriously by anyone 
\cite{Wiesner83}. There are Bennett and 
Brassard writing in 1984 about quantum key distribution and quantum bit 
commitment \cite{BB-84} --- the latter to be proved impossible a few years 
later, the former to be rediscovered by Ekert in 1991 for the benefit of 
physicists \cite{Ekert-91}. There is Peter Shor vindicating some previous 
speculations on quantum computing with a polynomial algorithm for 
factoring large integers \cite{Shor-94}. 

Before examining some of these topics, however, we pause to introduce the basic 
concepts underlying quantum information and computation.

\section{Key concepts underlying quantum information processing}
\label{sec:keyConcepts}
Quantum information processing examines the implications of replacing
our classical mechanically grounded notions of information and information
processing with quantum mechanically grounded ones. 
It encompasses quantum computing, quantum cryptography, quantum 
communication protocols, and beyond. 
The framework of quantum information processing has many similarities to 
classical information processing, but there are also several striking 
differences between the two. One difference is that the fundamental unit
of computation, the bit, is replaced with the quantum bit, or qubit. In this
section we define qubits and describe a few key properties of multiple
qubit systems that are central to differences between classical and quantum
information processing: the tensor product structure 
of quantum systems, entanglement, superposition, and quantum measurement.

\subsection{One qubit and its measurement}

The fundamental unit of quantum computation is the quantum bit,
or {\em qubit}. Just as there are many different physical implementations of a bit
(e.g. two voltage levels; toggle switch), there are many possible physical
implementations of a qubit (e.g. photon polarization; spin of an electron;
excited and ground state of an atom). And, just as in the classical case, information theory can abstract away from the specific physical instantiation and
discuss the key properties of qubits in an abstract way.

Classical bits are \textit{two-state} systems: the possible states are $0$ and $1$. Qubits are \textit{quantum two-level} systems: they can take infinitely many different states. However, crucially, \textit{different does not mean perfectly distinguishable} in quantum physics. In the case of qubits, any given state is perfectly distinguishable from one and only one other state: this is what makes the qubit the simplest, nontrivial quantum system. We cannot indulge here in explaining the mass of experimental evidence that lead physicists to accept such a counter-intuitive view, but we can sketch how one describes it mathematically.

Let us first pick a pair of perfectly distinguishable states and label them $\ket 0$ and $\ket 1$. These two states can be used to encode a classical bit: as a consequence, all of classical information and computation can be seen as special cases of the quantum ones. In order to describe the other possible quantum states, one has to use \textit{two dimensional complex vectors}:
\ban
\ket 0=\left(\begin{array}{c}1\\0\end{array}\right)\,,\;\ket 1=\left(\begin{array}{c}0\\1\end{array}\right) &\longrightarrow & \ket\psi= \left(\begin{array}{c}a\\ b\end{array}\right)\,=\,a\ket{0}+b\ket{1},
\ean
where $a$ and $b$ are complex numbers such that\footnote{
With this normalization and the convention that two vectors 
$\ket \psi$ and $e^{i\theta}\ket \psi$ differing by a global constant 
represent the same state, one is left with two real parameters. 
It can be shown that the possible states of a qubit are in one-to-one 
correspondence with the points on the unit sphere in 3 dimensions.} 
$\vert a\vert^2 + \vert b\vert^2 = 1$. One says that $\ket{\psi}$ is a \textit{superposition} of $\ket 0$ and $\ket 1$. Of course, superposition is relative to an arbitrary prior choice: physically, $\ket{\psi}$ is just as good a state as $\ket 0$ and $\ket 1$.
The relationship of distinguishability between two states $\ket \psi$ and $\ket \phi$ is captured by the absolute value of the scalar product, usually denoted $|\braket{\psi}{\phi}|$: the smaller this number, the more distinguishable the states are. In particular, perfectly distinguishable states are associated to orthogonal vectors.

The existence of infinitely many different states, but of finitely many (two, for qubits) perfectly distinguishable ones, is reflected in the readout of information a.k.a. \textit{measurement}. An elementary desideratum for measurement is that each outcome identifies the physical properties (i.e. the state). In quantum physics, therefore, a measurement consists in choosing a basis of orthogonal vectors as outcomes\footnote{More general notions of measurement have been defined, but they can all be put in the framework of a projective measurement (the type of measurement we have just defined) on a larger system.} so that \textit{after} the measurement, one can say with certainty in which state the system is. However, unless the state is promised to be already in one of the possible outcome states, the information about the state of the system \textit{prior} to the measurement is lost\footnote{A little bit of information on the state prior to the measurement is available: it could not have been orthogonal to the state that was detected, because otherwise it would have ended up in a different outcome for the measurement.}. 
In general thus, measurement creates a property rather than revealing it:
in other words, it is not possible to reliably measure an unknown state 
without disturbing it.

To give a concrete example, horizontal polarization $\ket{H}$ and
vertical polarization $\ket{V}$ 
of a photon can be perfectly distinguished. Similarly,
the two polarizations at $45^\circ$,
$\ket{\nearrow}=\frac{1}{\sqrt{2}}\ket{H}+\frac{1}{\sqrt{2}}\ket{V}$ and $\ket{\searrow}=\frac{1}{\sqrt{2}}\ket{H}-\frac{1}{\sqrt{2}}\ket{V}$
can be perfectly distinguished. 
The polarization of a photon can be measured in such
a way that the two outcomes are $\ket{H}$ and vertical $\ket{V}$,
or in such a way that the two outcomes are polarizations at $45^\circ$.
A photon with polarization $\ket{\nearrow}$, when measured in a way that
distinguishes horizontal polarization $\ket{H}$ from vertical $\ket{V}$ 
polarization, will become $\ket H$ with probability 
$|\braket{\nearrow}{H}|^2=1/2$ and $\ket V$ with 
probability $|\braket{\nearrow}{V}|^2=1/2$; but if it were measured 
instead with an apparatus that distinguished $\ket{\nearrow}$ from
$\ket{\searrow}$,
it would be found to be $\ket\nearrow$ with certainty.

\subsection{Uncertainty relations and no-cloning theorem}
\label{sec:uncertaintyPrin}

What we just discussed about measurement has implications that are worth spelling out, since they are well known and will form part of those quantum features which gives quantum key distribution and quantum computing their power.

First, we can come back to the idea of intrinsic randomness, seen from the angle of quantum measurement. We have said that in general the measurement prepares a given output state, but tells us little about an unknown input state. 
The reverse of this fact is also true: \textit{if the input state is 
known, the outcome of a measurement is in general random}. This must be 
the case in a theory in which there are infinitely many more states 
than those that can be perfectly discriminated: indeed, all the others 
cannot be perfectly discriminated and there will be an element of randomness. 
As we have seen previously, the violation of Bell's inequalties proves that this randomness is an intrinsic feature of our universe and not just a theoretical artefact.

From these observations it is easy to deduce the famous \textit{uncertainty relations}: given two measurements, most of the time they will project on different bases. Even if an input state gives a certain outcome for the first 
measurement, it will most probably give only statistical outcomes for 
the second. The original example of Heisenberg says that a particle 
cannot have both a well-defined position and a well-defined momentum.
While this example is particularly striking for its counter-intuitive 
character, one can define uncertainty relations even for a single qubit. 
Indeed, uncertainty relations are \textit{derived} notions and certainly not ``principles" as the wording often goes: they capture some specific quantitative aspects of the intrinsic randomness through statistical measures such as
variances or entropies.

Second, one can prove that it is \textit{impossible to copy reliably 
an unknown quantum state}, a statement that goes under the name 
of the \textit{no-cloning theorem}. This is obviously not the case with 
classical information: one can photocopy a letter without even reading 
its content. In the quantum case, for each set of fully distinguishable
states, there is a mechanism that can copy them. But the mechanism is
different for different sets. If the wrong mechanism is used, not only
does an incorrect copy emerge, but the original is also altered, 
probabilistically, in a way that is similar to the alteration by measurement.
If the state is unknown, it is impossible to reliably choose the
correct mechanism. 
The mathematical proof of the no-cloning theorem is simple 
(see Chapter 3 of \cite{sixpieces} or Chapter 5 of \cite{RPbook}). 
Here we can give a proof based on self-consistency of what we said. 
If it were possible to produce $N$ copies of a given state out of one, 
one could obtain sufficient statistical information to reconstruct the 
initial state: this procedure would discriminate all possible states, 
contrary to the assumption.


\subsection{Multiple qubit states: entanglement}

The use of vector spaces to describe physical systems has dramatic consequences when applied to the description of composite systems. 
To continue with the photon polarization example, consider two photons, propagating in different directions. Both photons can have horizontal polarization, and this we would write $\ket{H,H}$; or both vertical polarization, $\ket{V,V}$. But we are using a vector space: therefore, $a\ket{H,H}+b\ket{V,V}$ should be a possible state of the system too. Only, what does it describe?

Before sketching a reply, let us consider now the polarization of three 
photons: each photon is a qubit, which (upon choice of a basis) can be 
described by 2 complex parameters, so one would expect classically that 
three qubits are described by $3\times 2=6$ complex parameters. However, it is easy to convince oneself that an orthogonal basis consists of eight vectors, so in fact one needs $2^3=8$ complex parameters to describe three qubits\footnote{In the main text we use simplified counts. In fact, as we have seen in a previous footnote, two real parameters are sufficient to parametrize one qubit. Classically, therefore, one would expect that $3\times 2=6$ real parameters are necessary to represent three qubits. But the quantum count gives $2^3=8$ complex parameters, minus one real for the normalization, minus one real for a global phase factor, so in total 14 real parameters.}. The proof is simple: the eight states $\ket{H,H,H}$, $\ket{H,H,V}$, $\ket{H,V,H}$, $\ket{H,V,V}$, $\ket{V,H,H}$, $\ket{V,H,V}$, $\ket{V,V,H}$, $\ket{V,V,V}$ are perfectly distinguishable by measuring each photon in the $H-V$ basis; therefore, they must be described by mutually orthogonal vectors.

A direct generalization of this simple count shows that \textit{the
size of the state space grows exponentially with the number of components
in the quantum case}, as opposed to only linearly in the classical 
mechanical case. More specifically, the dimension of two complex 
vector spaces $M$ and $N$ when combined via the Cartesian product
is the sum of the two dimensions: $dim(M\times N) = dim(M) + dim(N)$.
The quantum formalism requires rather to combine systems using the \textit{tensor product}, in which case the dimensions multiply:
$dim(M\otimes N) = dim(M)dim(N)$, where $\otimes$ denotes
the tensor product.

In turn, this means that there are many states that do not have a 
simple classical counterpart. In particular, while the state of a 
classical system can be completely characterized by the state of 
each of its component pieces, most states of a quantum system cannot be
described in terms of the states of the system's components: 
such states are called {\it entangled states}. 
The two-photon superposition written above is an example of an 
entangled state for nonzero $a$ and $b$: $a\ket{H}\otimes\ket{H}+b\ket{V}\otimes\ket{V}\neq \ket{\psi}\otimes\ket{\phi}$. 
The statistics violating Bell's inequalities arise from measurements on such states.

In the following, we will be interested particularly in the case in which the system is a multiqubit system and the components of interest are the 
individual qubits. As it should be clear from the three photon example, 
a system of $n$ qubits has a state space of dimension $2^n$. 

\subsection{Sophisticated coda}

We want to add two remarks here, aimed at our more sophisticated readers. 
Other readers can simply skip this section.

In both quantum 
mechanics and classical mechanics, there are multiple meanings for the
word ``state.'' In classical mechanics, an attractive approach is
to take a state to mean a probability distribution over all possible
configurations rather than as the space of all configurations
as we are doing here. Similarly, in quantum mechanics, we have notions
of pure states and mixed states, with the mixed states being probability
distributions over pure states. We follow the convention that a state
means a pure state. In both cases, within the probabilistic framework,
 the pure states in the quantum mechanical case, or the set of 
configurations in the classical mechanical case, can be identified as 
the minimally uncertain, or maximal knowledge states. In the classical
case, such states have no uncertainty; they are the probability distributions
with value $1$ at one of the configurations and $0$ at all others. 
In the quantum mechanical case, the inherent uncertainty we discussed in
Section \ref{sec:Bell} means that even minimally uncertain states still
have uncertainty; while some measurements of minimally uncertain states
may give results with certainty, most measurements of such a state will
still have multiple outcomes. 

The tensor product structure in quantum mechanics also underlies
probability theory, and therefore appears in classical mechanics when
states are viewed as probability distributions over the set of configurations.
Unfortunately, the tensor product structure is not mentioned in
most basic accounts of probability theory even though
one of the sources of mistaken intuition about probabilities
is a tendency to try to impose the more familiar direct
product structure on what is actually a tensor product structure.
An uncorrelated distribution is the tensor product of its marginals. 
Correlated distributions cannot be reconstructed from their marginals.
For mixed quantum states, it is important to be able to distinguish
classical correlations from entanglement, which would require a more
sophisticated definition than the one we gave above. 
A major difference between classical probability theory 
is that minimally uncertainty states in the
classical case do not contain correlation, whereas in the quantum case,
most minimally uncertain states, the pure states, do contain entanglement. 
In other words, all minimally uncertain states in the classical setting
can be written as a tensor product of their marginals, whereas in the quantum
setting, most minimally uncertain states cannot be decomposed into
tensor factors. For more discussion of the relationship between classical
probability theory, classical mechanics, and quantum mechanics, see
\cite{Kuperberg, RPbook, Rieffel-07, WisemanMilburn}. For the remainder
of the discussion, we return to using ``state'' to mean ``pure state.''

\section{Quantum key distribution}

The best known application of quantum mechanics in a cryptographic 
setting, and one of the earliest examples of quantum information processing,
relates to the problem of establishing a key, a secret string of bits,
shared between two 
parties (usually called Alice and Bob) which they can use at some
later stage as the encryption key for sending secret messages between them. 
This problem is known as \textit{key distribution}. While the problem 
is easily solvable if Alice and Bob can meet, or if they share some 
secure channel over which they can communicate, it becomes much harder 
if all of their communication is potentially subject to eavesdropping.

In order to understand what quantum key distribution can and cannot
do for you, let us consider the classical scenario of a 
\textit{trusted courier}. Alice generates a string of bits, burns a copy 
of it in a DVD, and uses a courier to send it to Bob. Alice and Bob will 
then share the string of bits (the key), which will be private to them,
if everything went well. What could go wrong? One source of concern is 
the courier: he can read what is on the DVD, or allow someone else to read
it, while it is on its way. Quantum key distribution addresses 
this concern. As with any security protocol, there are threats that
a quantum key distribution protocol does not deal with. One is 
authentication. An eavesdropper Eve could convince the courier that
she is Bob, thus establishing a shared key between herself and Alice.
She can then set up a separate shared key between herself and Bob.
This scenario is called a man-in-the-middle attack. Conventional
means of authentication, through which Alice can be sure it was Bob who
received her string, exist and should be used with quantum key
distribution to guard against man-in-the-middle attacks.
Another concern is that someone might have tapped into Alice's private space 
(her office, her computer) while she was generating the string, 
or someone might tap in Bob's private space and read the key. 
If the private space of the authorized partners is compromised, 
there cannot be any security, and quantum information processing
cannot help. 

The crucial feature of quantum key distribution (QKD) is that, 
\textit{if the courier (a quantum communication channel) is corrupted 
by the intervention of Eve the eavesdropper, Alice and Bob will 
detect it}. Even more, they will be able to quantify how 
much information has leaked to Eve. On this basis, they can decide 
whether the string can be purified with suitable classical information 
processing and a key be extracted. If too much information has 
leaked, they will discard the entire string. At any rate, a 
non-secret key is never used. This eavesdropper-detecting functionality 
is inextricably linked to the no-cloning theorem, and as such could never 
be achieved using purely classical techniques.

\subsection{The physical origin of security}
\label{sec:physsec}

Quantum key distribution (QKD) is a huge research field, encompassing
a variety of different quantum key distribution protocols, error
correction and privacy amplification techniques, and implementation
efforts. Here we focus on the physical origin of security, that is, 
why Eve's intervention can be detected. 
For a basic introduction, see chapter 2 of \cite{sixpieces}; 
more experienced readers can consult a number of excellent review articles 
\cite{revqkd2,revqkd1,revqkd3}. 

The security of the first QKD protocol, proposed by Bennett and Brassard in 
1984~\cite{BB-84} and therefore called BB84, is based on the combination 
of the fact that \textit{measurement modifies the quantum state} and the 
fact that \textit{unknown quantum states cannot be copied} 
(the no-cloning theorem). Indeed, faced with her desire to learn 
what Alice is sending to Bob, 
Eve can try to look directly 
at the states or she can try to copy them to study at her leisure. 
Whether she tries to measure or copy, because the information
is encoded in a quantum state unknown to her, the measurement
or copying mechanism she chooses is almost certain to introduce
modification in the quantum state. Because of this modification,
Eve not only does not learn the correct state, but also does not
know what to send to Bob who is expecting to receive a state. Recall
that copying with the wrong mechanism disturbs the original as
well as the copy, so even in this case, she does not have an 
unmodified state to send along to Bob. The more Eve gets to know 
about the key, the more disturbance she causes in the 
state that reaches Bob. He and Alice can then compare notes
publicly on just some of the states he has received to check for 
modifications and thus detect Eve's interference. 

In 1991, Ekert re-discovered QKD~\cite{Ekert-91} using ideas with 
which we are already familiar: entangled states and Bell's inequalities.
If Alice and Bob share entangled states that violate Bell's inequalities, 
they share strong correlations which they can use to obtain a joint key
by measuring these states.
Alice's outcome is random for anyone except Bob, and vice versa. 
In particular, Eve cannot know those outcomes. At the opposite extreme, 
suppose that Eve knows perfectly the outcomes of Alice and Bob. 
Then those outcomes are no longer random, and a consequence, 
they cannot violate a Bell inequality. In summary, Ekert's protocol 
exploits a trade-off between the amount of \textit{violation of a 
Bell inequality} and the information that Eve may have about the outcomes.

\subsection{Device-independent QKD}
\label{sec:devindep}

In a very short time, physicists realized that, in spite of 
many superficial differences, BB84 and the Ekert protocol are 
two versions of the same protocol. Fifteen years later, physicists
realized that the two protocols are deeply different after all! 
In order to understand why, consider a security concern we have not
looked at yet. Where are Alice and Bob getting their devices,
the ones that create and detect the quantum states that are
used in the protocol?
Unless they are building the devices themselves, how do they
know that the devices are working as advertised? 
How can they be sure that Eve has not built or modified the devices to
enable them to behave in a different way, a way she can attack? 
They can detect when Eve interferes with a transmission. Can they 
somehow also detect when something fishy is going on with the devices
in the first place?

For BB84, it turns out that a high level of trust in the behavior 
of the apparatuses is mandatory. In particular, the protocol is 
secure only if the quantum information is encoded in a \textit{qubit}, 
that is a degree of freedom which has only two distinguishable states. 
If one cannot ensure this, the protocol is insecure: two classical bits 
suffice to simulate a ``perfect" run of the protocol \cite{acin2006bell}. 
Ekert's protocol, on the other hand, is based on Bell's inequalities. 
Referring back to what we wrote above, we see that this criterion is 
based only on conditional statistics: we have described it without 
having to specify either physical systems (photons, atoms...) or 
their relevant degrees of freedom (polarization, spin, ...). 
In short, one says that the violation of Bell inequalities is 
a \textit{device-independent} test \cite{acin2007}.

A few remarks are necessary. First of all, we are saying that 
anyone who can carry out Ekert's protocol can check
whether Bell's inequalities are violated. Creating devices that
exhibit violations of Bell's inequalities is much more complicated 
than simply testing for violations:
experimentalists, or the producers of QKD apparatuses, must know 
very well what they are doing. Second, even after establishing 
violations of Bell's inequalities, 
Alice and Bob cannot trust their devices blindly: an adversarial provider, 
for instance, might have inserted a radio which sends out the results of 
the measurements, a sort of Trojan horse in the private space. 
This is not a limitation of QKD alone: for any cryptographic protocol,
one must trust that there is no radio in any of the devices Alice 
and Bob are using in their private space. If Alice and Bob's 
measurement events are spacelike separated, meaning that no signal 
could travel between them during the time frame of the measurement process, 
they can be certain that the keys generated are truly random, provided 
they convincingly violate Bell's inequalities, even if their devices 
do contain radios or similar means of surreptitiously transmitting 
information. However if this communication continues after the key 
is generated, there is little to stop their devices betraying them and 
transmitting the key to Eve. Third, if Alice and Bob know quantum 
physics and find out that their devices are processing qubits, 
the Ekert protocol becomes equivalent to BB84.

With this understanding, we can go back to our initial topic of randomness 
and somehow close the loop, before focusing on quantum computation proper.

\subsection{Back to randomness: device-independent certification}
\label{sec:certifiedrandom}

Even to the one of us who was directly involved in the process, it is 
a mystery why the possibility of device-independent assessment was noticed 
only around 2006. Once discovered in the context of QKD, though, 
it became clear that the notion can be used in other tasks --- for instance, 
certified randomness generation \cite{pironio2010,colbeck2011}. 
Above, we discussed how Bell inequalities can convince anyone of the existence 
of intrinsic randomness in our universe. Rephrase it all in an 
industrial context, and one can conclude that the violation of 
Bell's inequalities can be used to \textit{certify randomness}.

This Bell-based certification has a unique feature: it guarantees 
that the random numbers are being produced on the spot, by the 
process itself. Let us explain this important feature in some detail. 
Consider first the usual statistical tests of randomness, which check for 
patterns in the produced string. Take a string that passes such a test 
and copy it on a DVD, then run the test on the copy. Obviously, the copy 
will pass the test too. Suppose you want to obtain a random string from
an untrusted source. How can you check that the string you receive is
random? As one example, how do you know the source is not sending the same
string to other customers? Classically, there is no way to check: a copy 
looks just as random as the original. Quantum mechanics does not provide
any additional ability to check for randomness after a string has been 
obtained, but a string of measurement outcomes from a Bell's experiment
that violates Bell's inequalities cannot have been preprogrammed at
the source, guaranteeing that the randomness is newly generated and not
a copy of a previously generated string. Again making use of the
free choice of settings in the Bell test, it is not possible to
use a predetermined string, no matter how random, to specify outcomes 
in a Bell test and still pass the test. Because the choice of measurement settings
changes unpredictably from test to test, it does not matter whether
the string passes classical tests for randomness or even came from a previous 
Bell test. What the analysis of violation of Bell's inequalities 
guarantees is that any predictable strategy for determining outcomes, 
even strategies  making use of strings certified as random by a previous 
Bell test, cannot produce outcomes that violate Bell's inequalities.
The combination of quantum physics with a test 
that contains an element of ``freedom" is what ultimately allows 
us to certify randomness as it is generated in this unique way.

\section{Quantum computing}
\label{sec:Qcomp}

The field of quantum computation examines how grounding computation
in quantum rather than classical mechanics changes how efficiently
computations of various types can be performed. We first present 
basic notions of quantum computation and then briefly discuss early 
algorithms before discussing the two most famous quantum algorithms, 
Shor's factoring algorithm and Grover's search algorithm. After a
brief discussion of quantum computational simulations of quantum systems,
we conclude the section with a discussion of known limitations on
quantum computation.

\subsection{Quantum computing basics}

We start by clarifying what a quantum computer is {\em not}. Just because
a computer makes use of quantum mechanical effects does not mean it is
a quantum computer. All modern computers make use of quantum mechanical
effects, but they continue to represent information as bits and act on the
bits with the same logical operations earlier machines used. The
physical way in which the logical operations are carried out may be 
different, but the logical operations themselves are the same. 

Quantum computers process qubits, and process them using quantum
logic operations, generalizations of classical logic operations that
enable, for instance, creation of entanglement between qubits.
Mirroring the situation with classical computation, any quantum 
computation can be broken down into a series of basic quantum logic gates. 
Indeed, any quantum mechanical transformation of an $n$ qubit system 
can be obtained by performing a sequence of one and two qubit operations. 
Unfortunately, most transformations cannot be performed efficiently in 
this manner, and many of the transformations which can be efficiently 
performed have no obvious use. Figuring out an efficient sequence of 
quantum transformations that can solve a useful problem is a hard problem 
and lies at the heart of quantum algorithm design.

Quantum gates act on quantum states, which means that they can act
on superpositions of classical values. Just as a single qubit can be
put in a superposition of the two distinguished states corresponding 
to bit values $0$ and $1$, a set of $n$ qubits can be placed in a 
superposition of all $2^n$ possible values of an $n$-bit string:
$0\dots00, 0\dots 01, \dots, 1\dots11$. Quantum circuits, made
up of quantum logic gates, can be applied to such a superposition. 
For every efficiently computable function $f$, there is an efficient
quantum circuit that carries out the computation of $f$. When applied
to a superposition of all of the $2^n$ possible input strings, this
circuit produces a superposition of all possible input/output pairs for $f$.
Such an application of a circuit to all possible classical inputs is
called {\it quantum parallelism}. Even though properties of quantum
measurement mean that only one of these input/output pairs can be
obtained from the superposition of all input/output pairs, the idea of 
``computing over all possible values at once'' is the most frequent
reason given in the popular press for the effectiveness of quantum 
computation. We discuss in Section \ref{sec:QParallel} further reasons 
why this explanation is misleading.

In this section, we touch on early quantum algorithms, Shor's algorithm and Grover's algorithm, as well as the
simulation of quantum systems, the earliest recognized application of
quantum computing. After the discovery of Grover's algorithm, there was
a five year hiatus before a significantly new quantum algorithm was
discovered. Not only have a variety of new algorithms emerged since then,
but also powerful new approaches to quantum algorithm design including
those based on quantum random walks, adiabatic quantum computation, 
topological quantum computation,
and one-way or measurement based computation which we will touch on
in Section \ref{sec:blindComp}. For a popular account of more recent
algorithms see \cite{Bacon10}. References \cite{Mosca08} and \cite{Childs10} 
provide more technical surveys. 

\subsection{Early quantum algorithms}

An early result in quantum computation showed that any classical algorithm
could be turned into a quantum computation of roughly equivalent complexity.
In fact, any reversible classical algorithm can be translated directly
into a quantum mechanical one. Any classical computation taking time $t$ 
and space $s$ can be turned into a reversible one with at most the slight 
penalty of $O(t^{1+\epsilon})$ time\footnote{In the discussion that follows we will make use of big-O notation, a standard way of characterizing the performance of an algorithm in theoretical computer science. In this notation, we say that a function $g(x)$ is in $O(f(x))$, if and only if for some sufficiently large constant $c$, $g(x)$ is bounded from above by $cf(x)$ for all values of $x$.}
and $O(s\log t)$ space \cite{Bennett-89}.
A classical deterministic computation that returns a result with certainty
becomes a deterministic quantum computation that also returns a result
with certainty. 
This fact provides another example of certainty in quantum mechanics
as previously discussed in Section \ref{sec:fewerWorlds}.
Quantum algorithms can 
be probabilistic, or they can be deterministic, returning a single final
result with probably $1$. Just to re-emphasize the earlier point that
quantum mechanics does not imply that ``everything happens," the obvious
deduction from the fact that an algorithm returns one result with certainty
is that the other results do not happen at all.

The early 1990s saw the first
truly quantum algorithms, algorithms with no classical analog 
that were provably better than any possible classical algorithm. 
The first of these, Deutsch's algorithm, was later generalized to the
Deutsch-Jozsa algorithm \cite{Deutsch-Jozsa-91}. 
These initial quantum algorithms were able to
solve problems efficiently with certainty 
that classical techniques can solve efficiently only with  
high probability. Such a result is of no practical interest since 
any machine has imperfections so can only solve problems
with high probability. Furthermore, the problems solved were
highly artificial. Nevertheless,
such results were of high theoretical interest since they proved 
that quantum computation is theoretically more 
powerful than classical computation. 

\subsection{Shor's factoring algorithm and generalizations}

These early results inspired Peter Shor's successful search for a 
polynomial-time quantum algorithm for factoring integers,
a well-studied problem of practical interest.
A classical polynomial-time solution has long eluded researchers.
Many security protocols base their security entirely on the computational
intractability of this problem. 
At the same time Shor discovered his factoring algorithm, he also
found a polynomial time solution for the discrete logarithm 
problem, a problem related to factoring that is also heavily used 
in cryptography. Shor's factoring and discrete log algorithms mean that
once scalable quantum computers can be built, all public
key encryption algorithms currently in practical use, such as RSA, will be 
completely insecure regardless of key length. 

Shor's results sparked interest in the field, but
doubts as to its practical significance remained. Quantum
systems are notoriously fragile. Key quantum properties, such as 
entanglement, are easily disturbed by environmental
influences. 
Properties of quantum mechanics, such as the no-cloning principle,
which made a straightforward extension of classical error correction
techniques based on replication impossible, made many fear that
error correction techniques for quantum computation would 
never be found. 
For these reasons, it seemed unlikely that reliable quantum 
computers could be built.
Luckily, in spite of widespread doubts as to whether
quantum information processing could ever be made practical, the theory
itself proved so tantalizing that researchers continued to explore it.
In 1996 Shor and Calderbank, and independently Steane, discovered
quantum error correction techniques that, in John Preskill's words 
\cite{Preskill98b}, ``fight entanglement with entanglement."
Today, quantum error correction is arguably the most mature
area of quantum information processing.

Both factoring and the discrete logarithm problem are 
{\it hidden subgroup problems} \cite{Lomont04}. In particular,
they are both examples of abelian hidden subgroup problems. Shor's
techniques easily extend to all abelian hidden subgroup
problems and a variety of hidden subgroup problems over 
groups that are almost abelian.
Two cases of the non-abelian hidden subgroup problem have received a lot
of attention: the symmetric group $S_n$ (the full permutation group
of $n$ elements) and the dihedral group $D_n$ (the group of symmetries
of a regular $n$-sided polygon). But efficient algorithms have eluded
researchers so far.
A solution to the hidden subgroup problem over $S_n$ would yield 
a solution to graph isomorphism, a problem conjectured to be NP-intermediate
along with factoring and the discrete log problem.
In 2002, Regev showed that an efficient algorithm to the 
dihedral hidden subgroup problem using Fourier sampling, a generalization
of Shor's techniques, would yield an efficient algorithm for the gap
shortest vector problem \cite{Regev02}. In 2003, Kuperberg found a
subexponential (but still superpolynomial) algorithm for the dihedral group
\cite{Kuperberg03} which he has recently improved \cite{Kuperberg11}.
Public key cryptographic schemes based on
shortest vector problems are among the most promising approaches
to finding practical public key cryptographic schemes that are 
secure against quantum computers.  

Efficient algorithms have been obtained for some related problems.
Hallgren found an efficient quantum algorithm for solving 
Pell's equation \cite{Hallgren-02}. Pell's equation, believed to be 
harder than factoring and the discrete logarithm problem, was the 
security basis for Buchmann-Williams key exchange and public key 
cryptosystems \cite{BuchmannWilliams88}. Thus, Buchmann-Williams joins 
the many public key cryptosystems known to be insecure in a world with 
quantum computers.
Van Dam, Hallgren, and Ip \cite{vanDam-03} found an efficient 
quantum algorithm for the shifted Legendre symbol problem, which means
that quantum computers can break certain algebraically homomorphic
cryptosystems and can predict certain pseudo-random number generators.

\subsection{Grover's algorithm and generalizations}

Grover's search algorithm is the most famous quantum algorithm after
Shor's algorithm. It searches an unstructured list of $N$ items
 in $O(\sqrt N)$ time. The best possible classical algorithm
uses $O(N)$ time. This speed up is only polynomial but, unlike for Shor's
algorithm, it has been proven that Grover's algorithm outperforms any possible
classical approach.  Although Grover's original algorithm succeeds only
with high probability, variations that succeed with certainty are known;
Grover's algorithm is not inherently probabilistic.

Generalizations of Grover's algorithm apply to a more restricted class 
of problems than is generally realized. It is unfortunate that Grover 
used ``database" in the title of his 1997 paper \cite{Grover-97}. 
Databases are generally highly structured and can be 
searched rapidly classically. Because Grover's algorithm does not 
take advantage of structure in the data, it does not provide a square root speed up for searching such databases.
Childs et al. \cite{Childs-08} showed that quantum computation 
can give at most a constant factor improvement for searches of 
ordered data such as that of databases. 
As analysis of Grover's algorithm focuses on query complexity, 
counting only the number of times a database or function must
be queried in order to find a match rather than considering the 
computational complexity of the process, it is easy to fall into 
the trap of believing that it must necessarily have better gate complexity,
the number of gates required to carry out the computation. 
This is not always the case, however, since the gate complexity of the
 query operation potentially scales linearly in $N$, as is the case for 
a query of a disordered database. The gate complexity of this operation
negates the $O(\sqrt N)$ benefit of Grover's algorithm, reducing its 
applications still further, in that the speed up is obtained only for 
data that has a sufficiently fast generating function.

As a result of the above restrictions, Grover's algorithm is 
most useful in the context of constructing algorithms based on black box 
queries to some efficient function. Extensions of Grover's algorithm 
provide small speed ups for a variety of problems
including approximating the mean of a sequence and other statistics,
finding collisions in $r$-to-$1$ functions, string matching, 
and path integration.  Grover's algorithm has also
been generalized to arbitrary initial conditions, 
non-binary labelings, and nested searches.

\subsection{Simulation}

The earliest speculations regarding quantum computation were spurred
by the recognition that certain quantum systems could not be
simulated efficiently classically \cite{Manin80, Manin07, Feynman-96}.
Simulation of quantum systems is another major application of quantum 
computing, with small scale quantum simulations over the past decade
providing useful results \cite{brown10, yung12}.
Simulations run on special purpose quantum devices provide
applications of quantum information processing to fields ranging 
from chemistry, to biology, to material science. They also support 
the design and implementation of yet larger special purpose quantum 
devices, a process that ideally leads all the way to the building 
of scalable general purpose quantum computers.

Many quantum systems can be efficiently simulated classically. After all,
we live in a quantum world and have long been able to 
simulate a wide variety of natural 
phenomena. Some entangled quantum systems can be efficiently simulated
classically, while others cannot. 
Even on a universal quantum computer, there are limits to what information can
be gained from a simulation. Some quantities, like the energy spectra
of certain systems, are exponential in quantity, so no algorithm, 
classical or quantum, can output them efficiently. 
Algorithmic advances in quantum simulation continue, while the 
question of which quantum systems can be efficiently
simulated classically remains open. New approaches to classical
simulation of quantum systems continue to be developed, many benefiting
from the quantum information processing viewpoint.
The quantum information processing viewpoint has led to improvements
in commonly used classical approaches to simulating quantum systems, 
such as the density matrix renomalization (DMRG) approach 
\cite{verstraete2004density} and the related matrix 
product states (MPS) approach \cite{perez2007matrix}.

\subsection{Limitations of quantum computing}
\label{sec:limitations}

Some popular expositions suggest that quantum computers would enable
nearly all problems to be solved substantially more efficiently than 
is possible with classical computers. Such an impression is false.
For example, Beals et al. \cite{Beals-01} proved for a broad class of problems
that quantum computation can provide at most a polynomial speed up. 
Their results have been extended and other means of establishing lower bounds
have also been found, yielding yet more problems for which it is known that
quantum computers provide little or no speed up over classical computers.
A series of papers established that quantum computers can search ordered 
data at most a constant factor faster than classical computers, and that
this constant is small \cite{Childs-08}. 
Grover's search algorithm is known to be optimal in that it is not possible to
search an unstructured list of $N$ elements more rapidly than 
$O(\sqrt{N})$. Most researchers believe that quantum computers cannot solve $NP$-complete
problems in polynomial time, though there is currently no proof of this (a proof would 
imply $P\ne NP$, a long standing open problem in computer science).

Other results establish limits on what can be accomplished with 
specific quantum methods.
Grigni et al.~\cite{Grigni01} showed that for
most non-abelian groups and their subgroups, the standard Fourier sampling
method, used by Shor and successors, yields exponentially little
information about a hidden subgroup. 
Aaronson showed that quantum approaches could not be used
to efficiently solve collision problems \cite{Aaronson02}.
This result means there is no generic quantum attack
on cryptographic hash functions that treats the hash function 
as a black box. By this we mean an attack that does not exploit 
any structure of the mapping between input and output pairs present 
in the function. Shor's algorithms break some cryptographic hash functions,
and quantum attacks on others may still be discovered, but Aaronson's
result says that any attack must use specific properties of the
hash function under consideration.

\section{Quantum information processing more generally}

Quantum information processing is a broad field that encompasses, for example, 
quantum communication and quantum games as well as quantum cryptography
and quantum computing. Furthermore, while quantum key distribution
is the best known quantum cryptographic protocol, many other types 
of protocols are known,
and quantum cryptography remains an active area of research. Here, we
briefly survey some quantum cryptographic protocols and touch on their
relation to quantum communication and quantum games. We then delve more
deeply into blind quantum computation, a recent discovery that combines 
cryptography and quantum computation.

\subsection{Quantum cryptography beyond key distribution}

While ``quantum cryptography'' is often used as a synonym for
``quantum key distribution,'' quantum approaches to a wide variety of other
cryptographic tasks have
been developed.  Some of these protocols 
use quantum means to secure
classical information. Others secure quantum information. Many
are ``unconditionally" secure in that their security is based 
entirely on properties of quantum mechanics. Others are only
quantum computationally secure in that their security depends on 
a problem being computationally intractable for quantum
computers. For example, while ``unconditionally" secure bit
commitment is known to be impossible to achieve through either
classical or quantum means, quantum computationally
secure bit commitments schemes exist as long as there are  
quantum one-way functions \cite{dumais00}.

Closely related to quantum key distribution schemes are protocols for 
unclonable encryption \cite{Gottesman-03}, a symmetric
key encryption scheme that guarantees that an eavesdropper cannot
copy an encrypted message without being detected. Unclonable
encryption has strong ties with quantum authentication. 
One type of authentication is digital signatures. Quantum digital 
signature schemes have been developed \cite{Gottesmann-Chuang-00},
but the keys can be used only a limited number of times. 
In this respect they resemble classical schemes such as 
Merkle's one-time signature scheme. 

Cleve et al. provide quantum protocols for $(k,n)$ threshold quantum 
secrets \cite{Cleve99}. Gottesman \cite{Gottesman99}
provides protocols for more general quantum secret sharing. 
Quantum multiparty function evaluation schemes
exist \cite{Crepeau-et.al-2002, Hayden04}. Brassard et al.~have shown 
that quantum mechanics allows for perfectly secure anonymous 
communication \cite{brassard2007anonymous}. Fingerprinting enables the 
equality of two strings to be determined efficiently with high probability by 
comparing their respective fingerprints \cite{ambainis-2003, Buhrman-01}. 
Classical fingerprints for $n$ bit strings need to be at 
least of length $O(\sqrt{n})$.  Buhrman et al. \cite{Buhrman-01} show 
that a quantum fingerprint of classical data can be exponentially 
smaller.

In 2005, Watrous showed that many classical
zero knowledge interactive protocols are zero knowledge against a 
quantum adversary \cite{Watrous05}. 
Generally, statistical zero knowledge protocols
are based on candidate NP-intermediate problems,
another reason why zero knowledge protocols
are of interest for quantum computation.
There is a close connection between quantum interactive protocols
and quantum games. Early work by Eisert et al. \cite{Eisert99} includes a 
discussion of a quantum version of the prisoner's dilemma. 
Meyer has written lively papers discussing other quantum games \cite{Meyer-game}.

\subsection{Blind quantum computation}
\label{sec:blindComp}
One area that combines both cryptography and computation is blind
quantum computation~\cite{broadbent2009universal}. Blind computation 
protocols address a situation 
that is becoming increasingly common with the advent of cloud computing, 
namely how to perform a computation on a powerful remote server in 
such a way that a person performing the remote computation, the 
client, can be confident that only she knows which computation was 
performed (i.e.~only she should know the input, output, and algorithm). 
While classical cryptographic techniques suffice in practice to prevent 
an eavesdropper from learning the computation if she can only access the 
communication between the client and the server, this security falls away 
if the eavesdropper has access to the server. In the case of blind quantum 
computation, the remote server is considered to be a fully fledged 
quantum computer, while the client is considered to have access only 
to classical computation, and the ability to prepare certain single qubit 
states.

This may seem like rather an odd task to focus on, but in a world where we
are digitizing our most sensitive information, maintaining the secrecy 
of sensitive material is more important than ever. Time on supercomputers 
is often rented, and so it is essentially impossible to ensure that nobody 
has interfered with the system. The problem becomes even more acute 
when we consider quantum computers, which will likely appear initially in 
only very limited numbers.

In 2001, Raussendorf and Briegel proposed a revolutionary new way of performing 
computation with quantum systems~\cite{raussendorf2001one}. Rather than using physical interactions 
between the qubits which make up such a computer to perform computation, 
as had been done up to that point, they proposed using specially 
chosen measurements to drive the computation. If the system was 
initially prepared in a special state, these measurements could be used 
to implement the basic logic gates that are the fundamental building 
blocks of any computation. This model of computation is purely quantum: 
it is impossible to construct a measurement-based computer according 
to classical physics. 

Measurement-based quantum computation supplements classical computation
with measurements on a special type of entangled state. The entangled state
is universal in that it does not depend on the type of computation being
performed. The desired computation is what determines the measurement sequence.
The measurements have a time ordering and the exact measurements to 
be performed depend on the results of previous measurements. 
Classical computation is used to determine what measurement should be  
performed next given the measurement results up until that point and to interpret
the measurement results to determine the final output of the computation, 
the answer to the computational problem. The measurements required are very
simple: only one qubit is measured at a time. One effect of this restriction
to single qubit measurements, as opposed to joint measurements of multiple
qubits, is that throughout the computation the entanglement can only 
decrease not increase, resulting in an irreversible operation. 
For this reason it is sometimes called 
``one way'' quantum computation. Measurement-based quantum computation
utilizing the correct sort of entangled state has been shown to be
computationally equivalent in power to the standard model of
quantum computation, the circuit model. The outcomes of the measurements,
given that they are measurements on an entangled state, exhibit a high
degree of randomness. It is a surprising and elegant result that these
random measurement outcomes add sufficient power to classical computation
that it becomes equivalent in power to quantum computation. 

Measurement-based quantum computation provides a particularly 
clean separation between the classical and quantum parts of a quantum
algorithm. It also suggest a fundamental connection between entanglement
and the reason for the power of quantum computation. But the issues
here are subtler than one might expect at first. In 2009, two groups of
researchers \cite{Bremner09,Gross09} showed that if a state is 
too highly entangled it cannot support quantum computation. Specifically, 
Gross et al.~showed that if the state is too highly entangled, the outcomes
of any sequence of measurements can be replaced by random classical
coin flips \cite{Gross09}. Thus, if a state is too highly entangled, 
the resulting outcomes are too random to provide a quantum resource. We will 
return to this point later when we discuss the mystery surrounding
the sources of quantum computing's power. As Gross et al.~conclude 
with respect to entanglement, ``As with most good things, 
it is best consumed in moderation."

This new model of measurement-based quantum computation opens many promising routes for building large scale quantum computers. Indeed, many researchers are currently working on architectures for distributed quantum computers based on this model which may lead to large scale quantum computers. However, measurement-based computation is not simply a way to build better computers, but rather a new way to think about computation.

In particular, measurement-based quantum computation provides a 
convenient lens with which to examine whether 
or not it is possible to perform a blind computation on a remote computer. 
The uncertainty principle allows for more information to 
be encoded in a quantum state than can be accessed through measurements. 
As measurement-based computation allows quantum computation to be 
constructed from measurements on quantum states together with a classical 
rule for adapting subsequent measurements, by using subtly different 
initial quantum states for the computation, different logic gates can 
be implemented. Each possible initial state is chosen in such a way that 
they yield identical results for any possible measurement made by the server, 
but yet each nudges the computation in a different direction. As a result, 
it is possible to perform arbitrary calculations blindly.

In fact, quantum properties enable us to take the security
one step further. By adapting standard techniques 
to detect errors in quantum computers, it is possible to 
detect any interference with the blind computation 
\cite{fitzsimons2012unconditionally}. Taken together, these 
results provide us with a way to ensure that our computation 
remains private and correct without needing to trust the computer or 
those who have access to it.

Abadi, Feigenbaum, and Kilian \cite{Abadi89} showed that, only in 
the unlikely event that a famous conjecture in complexity theory fails, 
information theoretically secure blind computation cannot be carried out on 
a classical computer. If only computational security is required,
classical solutions are possible, though it was only in 2009 that
the first such scheme was found, Gentry's famous fully homomorphic encryption
scheme~\cite{gentry2009fully}.
Fully homomorphic encryption is most commonly described as enabling
universal computation on encrypted data by a party, say a server, that 
does not have access to a decryption key and learns nothing about the 
encrypted data values. The server returns the result of the computation
on the encrypted data to a party who can decrypt it to obtain meaningful
information. Fully homomorphic encryption can also be used to hide
the computation being carried out, thus achieving a form of blind 
computation, but with only a computational security guarantee.

However, there are significant differences between the capabilities of
blind quantum computation and classical fully homomorphic encryption. Most importantly, blind quantum computation allows the client to boost their computational power to an entirely different computational complexity class (from P to BQP), unlike known homomorphic encryption schemes. 
Further, a blind quantum computation can be authenticated, enabling the detection of any 
deviation from the prescribed computation with overwhelming probability. 
The security provided by the protocols is also different: while known 
fully homomorphic encryption schemes rely on computational assumptions 
for their security, the security of the blind computation protocol 
described in \cite{broadbent2009universal} can be rigorously proved 
on information theoretic grounds.

\section{Classical lessons from quantum information processing}

The quantum information processing viewpoint provides insight
into complexity issues in classical computer science and
has yielded novel classical algorithmic results and methods.
The usefulness of the complex perspective for evaluating
real valued integrals is often used as an analogy to explain
this phenomenon.
Classical algorithmic results stemming from the insights of quantum
information processing include lower bounds for problems involving 
locally decodable codes, local search, lattices, reversible circuits, 
and matrix rigidity. Drucker and de Wolf \cite{Drucker09} survey a 
wealth of purely classical computational
results, in such diverse fields as polynomial approximations, matrix
theory, and computational complexity, that resulted from taking
a quantum computational view.  

In two cases, quantum arguments have been used to establish security
guarantees for purely classical cryptographic protocols.
Cryptographic protocols usually rely on the empirical hardness of
a problem for their security; it is rare to be able to prove
complete, information theoretic security. When a cryptographic protocol
is designed based on a new problem, the
difficulty of the problem must be established before the security
of the protocol can be understood. Empirical testing of a problem
takes a long time. Instead, whenever possible,
``reduction" proofs are given that show that if the new problem were
solved it would imply a solution to a known hard problem.
Regev designed a novel, purely classical cryptographic
system based on a certain lattice problem \cite{Regev-05}. 
He was able to reduce a known hard problem to this problem, but 
only by using a quantum step as part of the reduction proof.
Gentry, for his celebrated fully homomorphic
encryption scheme \cite{gentry2009fully}, provides multiple reductions, 
one of which requires a quantum step.

\section{Implementation efforts}

Over the past two decades since the discovery of Shor's and Grover's 
algorithms, progress in realizing a scalable quantum computer has begun
to gather pace. Technologies based on liquid state nuclear magnetic 
resonance techniques (NMR) provided a test bed for many proof of concept 
implementations of quantum algorithms and other quantum information 
processing tasks. However, because of problems cooling, liquid state NMR 
is not considered a viable route to a scalable quantum computer. 
The leading candidates for viable routes to scalable quantum computers have
long been ion trap and optical quantum computing. 
Recently, however, progress in superconducting qubits has shown 
significant promise for scalable quantum computing. 
Superconducting quantum processors could be constructed using techniques 
and facilities similar to today's semi-conductor based processors. 
Recently IBM has demonstrated gate fidelities approaching the threshold 
necessary for fault-tolerant quantum computation~\cite{chow2012complete}.

While scalable quantum computing has not yet been achieved, quantum key 
distribution has already been developed into a viable technology. Today, 
commercial quantum key distribution systems are already available from a 
number of manufacturers including id Quantique and MagiQ Technologies. 
Other quantum cryptographic techniques have not yet matured to this level, 
but many, including blind quantum computation \cite{barz2012demonstration}, 
have been demonstrated in a laboratory setting.

\section{Where does the power of quantum computing come from?}

In contrast to the case for quantum key distribution, the source of
the power of quantum computation remains elusive. Here we review some of the
explanations commonly given, explaining both the limitations of and
the insights provided by each explanation.

\subsection{Quantum parallelism?}
\label{sec:QParallel}

As discussed in Section \ref{sec:Qcomp},
the most common reason given in the popular press for the power
of quantum computation is ``quantum parallelism.''
However, quantum parallelism is less powerful than it may initially appear.
We only gain information by measuring, but
measuring results in a single input/output pair, and a random one at
that. By itself, quantum parallelism is useless. 
This limitation leaves open the possibility that quantum parallelism
can help in cases where only a single output, or a small number of outputs,
is desired. While it suggests a potential exponential speed up for
all such problems, as we saw in Section \ref{sec:limitations}, 
for many problems it is known that no such speed up is possible.

Certain quantum algorithms that were initially phrased in terms
of quantum parallelism, when viewed in a clearer light, have little
to do with quantum parallelism. 
Mermin's explanation of the Bernstein-Vazirani algorithm, originally
published in his paper {\it Copenhagen Computation: How I Learned to 
Stop Worrying and Love Bohr} \cite{Mermin04}, contributed
to this enlightenment. He was the first to see that, without changing
the algorithm at all, just viewing it in a different light, the 
algorithm goes from one phrased in terms of quantum parallelism 
in which a calculation is needed to see that
it gives the desired result, to one in which the outcome is evident.
The Bernstein-Vazirani algorithm \cite{Bernstein-Vazirani-97}, and 
Mermin's argument in particular, deserves to be better known because of 
the insight they give as to how best to view quantum computation.

\subsection{Exponential size of quantum state space?}

A second popular explanation is the exponential size of the
state space. This explanation is also flawed. To begin with, 
as we have seen, exponential spaces also arise in classical probability 
theory. Furthermore, what would it mean for an efficient algorithm to
take advantage of the exponential size of a space?
Even a superposition  of the exponentially many possible values
of an $n$-bit string is only a single state of the quantum state space.
The vast majority of states cannot even be approximated by an 
efficient quantum algorithm \cite{Knill-95}.
As an efficient quantum algorithm cannot even come close to most
states in the state space, quantum parallelism does not, and 
efficient quantum algorithms cannot, make use of the full state space.  

\subsection{Quantum Fourier transforms?}

Most quantum algorithms use quantum Fourier transforms (QFTs). The
Hadamard transformation, a QFT over
the group ${\bf Z}_2$, is frequently used
to create a superposition of $2^n$ input values.
In addition, the heart of most quantum algorithms makes use of 
QFTs. Shor and Grover both use QFTs.
Many researchers speculated that quantum
Fourier transforms were a key to the power of
quantum computation, so it came as a surprise when 
Aharonov et al.~\cite{Aharonov-06} showed that QFTs
are classically simulable. Given the ubiquity of quantum
Fourier transforms in quantum algorithms, researchers continue
to consider QFTs as one of the main tools
of quantum computation, but in themselves they are not sufficient.

As any quantum computation can be constructed out of a series of gates 
consisting of quantum Fourier transforms and transformations that
preserve the computational basis, it has been suggested that the 
minimum number of layers of Fourier transforms required for an 
efficient implementation of a particular quantum transformation 
gives rise to a hierarchy (known as the Fourier Hierarchy) containing 
an infinite number of levels, which cannot be collapsed while maintaining 
polynomial circuit size~\cite{shi2005quantum}. The zeroth and first levels 
of such a hierarchy correspond to the classical complexity classes P and BPP 
respectively, while many interesting quantum algorithms, such as Shor's 
factoring algorithm, occupy the second level. Nonetheless, the truth of 
the Fourier Hierarchy conjecture that the levels cannot be collapsed
remains an open problem.

\subsection{Entanglement?}

Jozsa and Linden \cite{Jozsa-Linden} show that any quantum algorithm 
involving only pure states that achieves exponential speed up over 
classical algorithms must entangle a large numbers of qubits. 
While entanglement is necessary for an exponential speed up,
the existence of entanglement is far from sufficient to guarantee
a speed up, and it may turn out that another property better
characterizes what enables a speed up. Many entangled systems have
been shown to be classically simulable \cite{Markov-05,Vidal-03}. 
Indeed, the Gottesman-Knill theorem~\cite{aaronson2004improved}, 
as well as results on the classical simulation of match 
gates~\cite{valiant2001quantum}, have shown that there exist 
non-classical computational models that allow for highly entangled 
states which are efficiently classically simulable.
Furthermore, if one looks at query complexity instead of algorithmic 
complexity, improvements can be obtained with no entanglement 
whatsoever. Meyer \cite{Meyer-00} shows that in the course of the
Bernstein-Vazirani algorithm, which achieves an $N$ to $1$
reduction in the number of queries required, no qubits become
entangled. Going beyond quantum computation it becomes more obvious
that entanglement is not required to reap benefits. For example, the 
BB84 quantum key 
distribution protocol makes no use of entanglement.
While measurement-based quantum computation, discussed in 
Section \ref{sec:blindComp}, graphically illustrates the use of
entanglement as a resource for quantum computation, it turns out
that if states are too highly entangled, they are useless for
measurement-based quantum computation \cite{Bremner09,Gross09}.
In the same paper in which they showed that entanglement is
necessary, Jozsa and Linden end their abstract with
``we argue that it is nevertheless misleading to view entanglement
as a key resource for quantum-computational power." \cite{Jozsa-Linden}.
The reasons for quantum information processing's power remains mysterious;
Vedral refers to ``the elusive source of quantum
effectiveness" \cite{Vedral09}.

\section{What if quantum mechanics is not correct?}

Physicists do not understand how to 
reconcile quantum mechanics with general relativity. A complete physical
theory would require modifications to general relativity, 
quantum mechanics, or both. Any modifications to quantum mechanics
would have to be subtle as the predictions of quantum mechanics 
hold to great accuracy, and most predictions of quantum mechanics will continue to  hold, 
at least approximately, once a more complete theory is found. Since no
one yet knows how to reconcile the two theories, no one knows what, if any,
modifications would be necessary, or whether they would affect the
feasibility or the power of quantum computation. 

Once the new 
physical theory is known, its computational power can be analyzed. 
In the meantime, theorists have
looked at what computational power would be possible if certain changes
in quantum mechanics were made. So far these changes imply greater
computational power rather than less.
Abrams and Lloyd \cite{AbramsLloyd98} showed that if quantum mechanics were 
non-linear, even slightly, 
all problems in the class $\#P$, a class that contains all NP problems
and more, would be solvable in polynomial time. Aaronson \cite{Aaronson04} 
showed that any change to one of the exponents in the axioms of 
quantum mechanics would yield polynomial time solutions to 
all PP problems, another class containing NP. These two results 
are closely related, in that a classical computer augmented with the power 
to solve either class of problems efficiently would have identical power. 
With these results in mind, Aaronson \cite{Aaronson08}
suggests that limits on computational power should 
be considered a fundamental principle guiding physical theories, much
like the laws of thermodynamics. 

\section{Conclusions}

We hope this glimpse of quantum information processing has intrigued you.
If so, there are many excellent resources for learning more, from
books on quantum computation \cite{NCbook,RPbook} to 
arxiv.org/archive/quant-ph where researchers post papers with their 
most recent results. 

Advances in quantum information processing are also driving the development 
of other technologies beyond computation and communication. Quantum 
information techniques have led to advances in lithography, providing a means
to affect material at scales below the classical wavelength 
limit~\cite{PhysRevLett.85.2733}. Quantum information processing has
motivated significant strides in our ability to control quantum 
systems~\cite{WisemanMilburn}. Further, quantum mechanics allows 
for significant improvements in the performance of a variety of sensors. 
Theoretical improvements have been demonstrated in a number of settings, 
initially restricted to simple parameter 
estimation~\cite{boixo2007generalized,giovannetti2004quantum,yurke1986input}, 
but later extended to imaging and other complex tasks~\cite{lloyd2008enhanced}.
Experimentally, such quantum techniques have been demonstrated to provide 
increased accuracy in estimating phase shifts induced by optical 
materials~\cite{nagata2007beating}, 
spectroscopy~\cite{leibfried2004toward,roos2006designer}, 
and in estimating magnetic field strenghts~\cite{jones2009magnetic}.

Many open problems remain. Some are of a fundamental nature. What
does nature allow us to compute efficiently? What does nature allow
us to make secure? Others are of a more practical nature. How will
we build scalable quantum computers? For what problems are there effective
quantum algorithms? How broad an impact will quantum information 
processing have? At the very least, quantum computation, and
quantum information processing more generally, has changed forever 
how humanity thinks about and works with physics, computation, and 
information.

\bibliographystyle{plain}
\bibliography{qc}

\end{document}